# Frequency characteristics of strip waveguides incorporating bulk lithium niobate crystals as a dielectric

Nickolay D. Malyutin [1], Artush A. Arutyunyan [1,*], Anton G. Loschilov [1], George A. Malyutin [1] and Dmitriy Minenko [1]

[1] Department of Design of Units and Components for Radioelectronic Systems, Tomsk State University of Control Systems and Radioelectronics, 634050 Tomsk, Russia;
* Correspondence: arutyunyan18@mail.ru

**Abstract:** The elements of the metamaterial made in the form of wave-guiding coplanar strip line type strip structures, the current-carrying strip of which is located on the upper surface of the substrate, have been experimentally investigated. On the same surface, side screens, connected to the lower grounded base, through metallized holes are made. A volumetric lithium niobate crystal is placed on the upper surface of the strip lines. Frequency dependences of the modulus and phase of the transmission coefficient and reflection coefficient of models with a lithium niobate bulk crystal are measured when a linear-frequency modulated signal in the range from 10 MHz to 25 GHz with a fast change of the incident wave direction is applied to the strip line input. The occurrence of quasi-chaotic oscillations in the tested models was demonstrated. Modeling of the strip structure allowed to determine the frequency dependence of the complex propagation coefficient, indicating the cause of quasi-chaotic oscillations due to the occurrence of a large number of normal waves and their interference. Frequency regions with negative group delay were detected.

## 1. Introduction

In [1] an assumption is made about the existence of a substance with simultaneously negative dielectric permittivity $\varepsilon < 0$ and magnetic permeability $\mu < 0$. Based on formal analysis, a conclusion is made, that in these substances the phase velocity is opposite to the energy flow. At the moment, materials with "double negative" $\varepsilon < 0$ and $\mu < 0$ have not been found in nature [2]. Therefore, the search for such materials is carried out in artificially created environments (chiral environments), also called metasurfaces or metamaterials [3–7]. Studies of the problem of existence of substances with $\varepsilon < 0$ and $\mu < 0$ on examples of artificially created quasi-continuous media have stimulated the creation of devices and systems with space-time modulation of parameters, non-reciprocal reflectors, metasurfaces for transformation of received and reflected signals for a wide variety of applications [8–21]. In [9], a theoretical description of the Pockels effect in crystals under static field control was proposed. Analytical expressions, determining phase velocities and polarization of plane light waves in a crystal in the main crystallographic directions under the influence of a static field were obtained. Nowadays, due to the rapid development of photonics and radiophotonics, crystals with unique properties are widely used to design acousto-optical converters, electro-optical modulators, delay lines, frequency-selective devices, etc. [22–24]. Of great interest is the study of wave properties of devices in which, for example, lithium niobate is used as a functional substrate. Thus, article [22] shown, that in addition to the surface acoustic wave modulating the optical signal, a significant energy of the electric RF signal can be used to excite the parasitic pseudo-PAV and delayed pseudo-PAV. Due to the proximity of the resonant excitation frequency of the delayed pseudo-PAV to the SAW excitation frequency, additional noise appears in the optical signal. In [23], studies of the characteristics of integrated electro-optic modulators based on lithium niobate crystals at low-frequency modulation up to 1 MHz are presented. In [24], a microwave slot-type design for a bulk light modulator is analyzed. The known method of equivalent wave impedances of waveguide structures in the form of a section of a rectangular waveguide filled with dielectric docked with a section of a hollow rectangular waveguide was applied. The frequency characteristics of the model under consideration are calculated. It is shown that at small dimensions of the slot height it is possible to achieve a sufficiently large broadband of the order of 1–2 GHz.

2. Module Designs

In this paper, we consider a metasurface element, made in the form of a waveguide strip structure of the coplanar strip line (CSL) type, the current-carrying strip of which is located on the upper surface of a foil dielectric substrate [25]. On the same surface side screens connected to the lower grounded base through metallized holes are made. A volumetric lithium niobate crystal in the form of a bar with the size L×W×H=39×18×29 mm with a tight fit to the surface of the current-carrying strip and side screens is installed on the upper surface of the strip line. The bottom surface of the substrates is metallized. The input and output of the strip line are provided with coaxial-strip transitions. The designs of the two studied models of strip waveguides are shown in Figure 1 a, b.

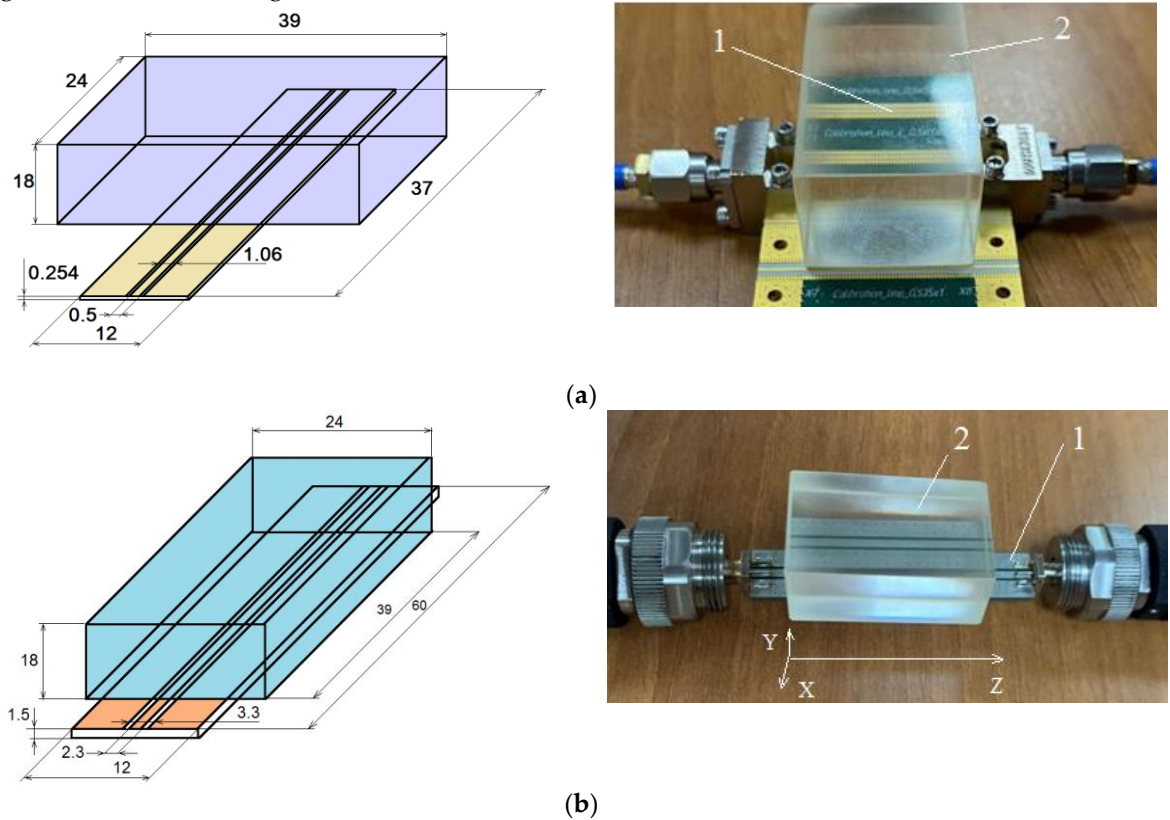

(a)

(b)

**Figure 1.** Model designs of strip waveguides in the form of coplanar strip lines 1 on dielectric substrate, on the upper plane of which a crystal of lithium niobate 2 is placed: a) model I, made on foil dielectric RO4350B, crystal location along the Z axis perpendicular to the current-carrying strip; b) model II, made on foil dielectric FR4, crystal location along the Z axis parallel to the current-carrying strip.

## 3. Results and Discussion

The task was set to study the frequency dependence of the transmission coefficient $S_{21}(f)$ and reflection coefficient $S_{11}(f)$ of strip structures, when filling the upper half-space with a nonlinear material with anisotropic piezoelectric and segmentoelectric properties, such as $LiNbO_3$ crystal. The features of the properties of the strip structure with the bulk $LiNbO_3$ crystal were studied when it is excited by the electromagnetic field of the strip line and an LFM signal is input in the range from 10 MHz to 25 GHz. The R4226 vector circuit analyzer was calibrated in dual-port mode, which means changing the direction of the incident wave to either the left coaxial-strip junction (input) or the right junction (output becoming the input). Both modules (Figure 1) are 50 Ω wave impedance matched in the absence of dielectric filling of the upper half-plane. Model I of the CPL (Figure 1a) is made of 0.254 mm thick foil dielectric RO4350B, relative permittivity $\varepsilon_r = 3.48$. Crystal Z axis was oriented as perpendicular to the current-carrying strip. Model II (Figure 1b) is made of 1.5 mm thick foil dielectric FR4. Crystal Z axis was oriented parallel to the current-carrying strip. The CPL design has two gaps between the current-carrying strip and the side shields. Undoubtedly, the basic physical regularities of light wave propagation in $LiNbO_3$ crystals, considered in [9] and other works, when exposed to microwave oscillations will be similar, but with differences due to the comparability of the length of propagating waves with the dimensions of the crystal. We have measured the frequency characteristics of models I, II, shown in Figure 1, in the frequency range from 10 MHz to 25 GHz on vector circuit analyzer R4226 produced by JSC "NPF "Mikran". The crystal was exposed to a linearly frequency-modulated field of a quasi-T-wave, the direction of which was changed by standard port switching. The modulus and phase of the transmission coefficient

$S_{21}(f)$ and the modulus and phase of the reflection coefficient $S_{11}(f)$ were measured. The measurement results are shown in Figure 2–Figure 4.

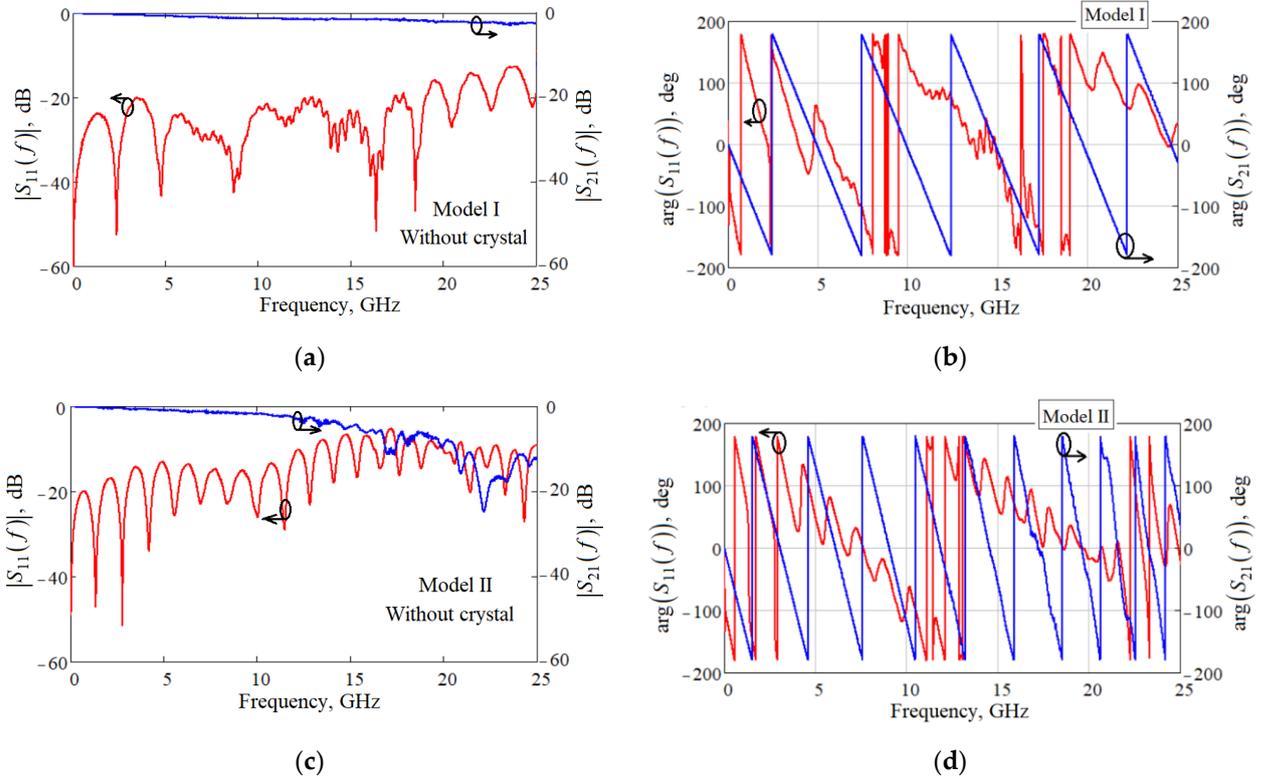

**Figure 2.** Frequency dependence of: *a*) transmission coefficient $|S_{21}(f)|$ and reflection coefficient $|S_{11}(f)|$ of the model I CPL without crystal; *b*) phase of the transmission coefficient $\varphi_{21}=\arg[S_{21}(f)]$ and phase of the reflection coefficient $\varphi_{11}=\arg[S_{11}(f)]$ of the model I CPL without crystal; *c*) $|S_{21}(f)|$ and $|S_{11}(f)|$ of the model II CPL without crystal; *d*) $\varphi_{21}=\arg[S_{21}(f)]$ and $\varphi_{11}=\arg[S_{11}(f)]$ of the model II CPL without crystal.

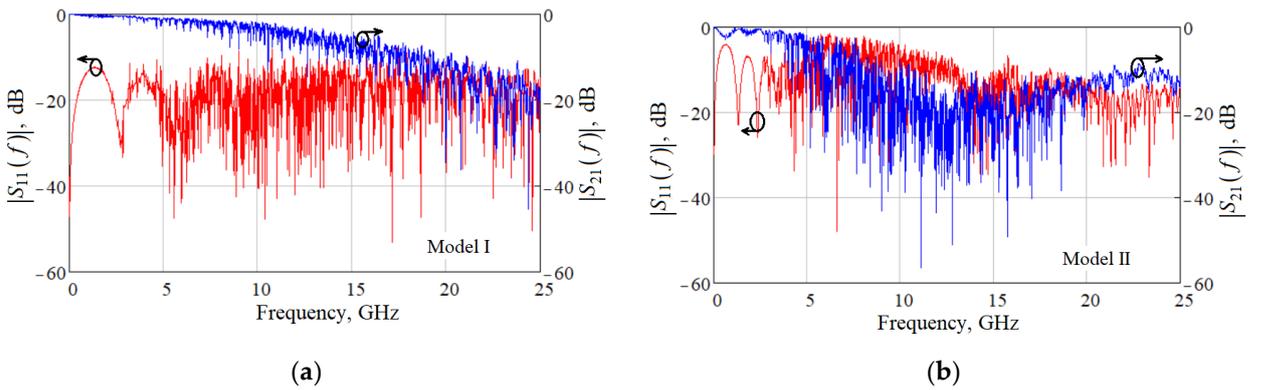

**Figure 3.** Frequency dependence of the transmission coefficient $|S_{21}(f)|$ and the reflection coefficient $|S_{11}(f)|$ (*a*) model I of CPL with lithium niobate crystal and (*b*) model II of CPL with lithium niobate crystal.

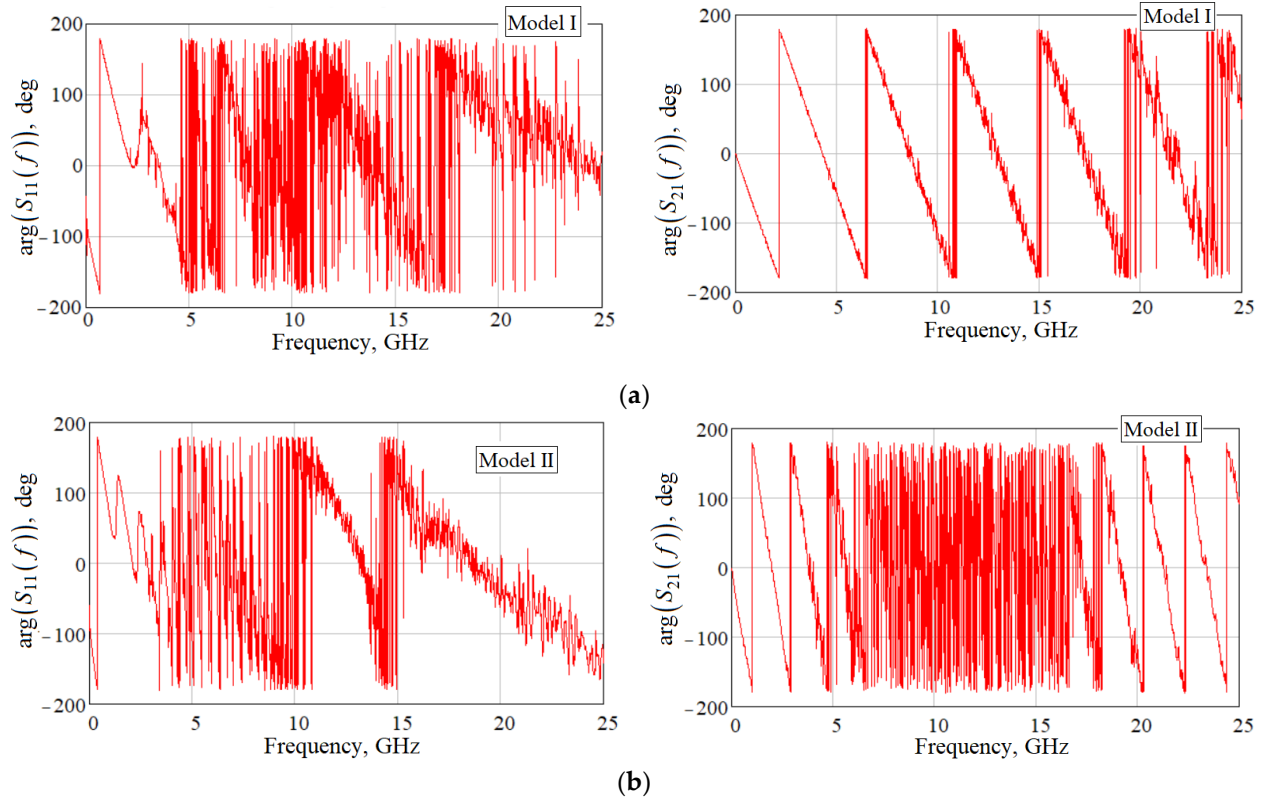

**Figure 4.** Frequency dependence of the phase of the transmission coefficient $\varphi_{21} = \arg[S_{21}(f)]$ and the phase of the reflection coefficient $\varphi_{11} = \arg[S_{11}(f)]$ (a) model I CPL with lithium niobate crystal and (b) model II CPL with lithium niobate crystal.

From the comparison of graphs $|S_{21}(f)|$, $\arg[S_{21}(f)]$, $|S_{11}(f)|$ and $\arg[S_{11}(f)]$ we see the occurrence of quasi-chaotic oscillations in the tested model at partial filling of the upper half-plane with lithium niobate crystal. The presence of quasi-chaotic oscillations can be explained by several reasons. The electromagnetic field in the CPL gaps excites electromagnetic oscillations in the lithium niobate crystal by two formed dipole-type sources in in-phase mode. The power flux in the coplanar line is conventionally divided into two coupled fluxes. One flow propagates in a dielectric substrate, the other in a crystal. The substrate is made of a homogeneous and isotropic dielectric, and the lithium niobate crystal, being uniaxial and characterized by a trigonal crystal system, has anisotropy of dielectric properties [8]. When a quasi-T-wave propagates along the current-carrying coplanar strip and, respectively, along the Z axis of the crystal, the electric field has components $E_X$, $E_Y$ along the X and Y axes [9]. This causes an electro-optical effect for waves polarized in the directions of the crystallographic axes. The velocities of the considered waves with components $E_X$, $E_Y$ are different because the relative dielectric permittivities $\varepsilon_{11} \neq \varepsilon_{33}$ are different [8, 9]. A "slow" wave with velocity $v_1$ and $v_2$ a "fast" wave are formed. Due to the large difference between $v_1$ and $v_2$, the conditions of eigenwave interference in the strip structure arise [10]. The presence of the crystal leads to an increase in the effective dielectric constant of the CPL segment, resulting in a drop in the wave impedance and an increase in the reflection coefficients $|S_{11}(f)|$ and $|S_{21}(f)|$. The crystal is three-dimensional, so starting from a frequency of about 5 GHz a very complex composition of natural waves appears. At the same time, a quasi-T wave transformation occurs in the current-carrying strip. This transformation is caused by the influence of the field components arising inside the crystal due to polarization and "splitting" of the quasi-T wave of the strip structure into: firstly, the crystal's own waves as a bulk dielectric resonator with a spectrum of its own waves; secondly, the waves associated with the main quasi-T wave, the phase velocities of which differ significantly. The appearance of combined waves due to the transformation of quasi-T-wave leads to their interference, since their phase velocities are different. The above processes lead to amplitude and phase quasi-chaotic oscillations $|S_{21}(f)|$, $\arg[S_{21}(f)]$, $|S_{11}(f)|$ and $\arg[S_{11}(f)]$, which are illustrated by the graphs of Figure 2–Figure 4. The study of model II also showed the occurrence of quasi-chaotic oscillations with larger amplitude and rate of change $|S_{21}(f)|$, $\arg[S_{21}(f)]$, $|S_{11}(f)|$ and $\arg[S_{11}(f)]$. The modeling of strip waveguides including bulk lithium niobate crystals as a dielectric was performed by analogy with [24] based on the method of equivalent primary and secondary parameters of the waveguide structure in the form of a coplanar strip line segment. Based on the obtained experimental dependences of the scattering matrix coefficients, we calculated the main primary equivalent parameters of the strip line when a quasi-T wave propagates in it: linear capacitance $C_0$ and linear inductance $L_0$, effective dielectric permittivity $\varepsilon_{eff}$. Secondary parameters have been determined: complex propagation coefficient $\gamma$, wave impedance $Z_0$. These and other parameters are strongly frequency dependent due to the variation of relative dielectric permittivity of $LiNbO_3$ crystal from frequency [8]. The representation of the investigated structures (Figure 1) in the form of coplanar strip line segments with frequency-dependent parameters showed almost complete agreement with the experimental characteristics, which indirectly confirms the discussed mechanism of quasi-chaotic oscillations formation in the considered model structures.

In the frequency dependences of $\arg[S_{21}(f)]$ and $\arg[S_{11}(f)]$ of models I and II, an increase of $\partial[\arg[S_{21,11}(f)]]/\partial f$ derivatives was observed in the wide frequency band above 5 GHz compared to the low-frequency band (Figure 3, Figure 4). To compare $\arg[S_{21}(f)]$ and $\arg[S_{11}(f)]$ of the studied models, the frequency dependence of the total phase $\psi(f)$ [26] was measured, which showed that at frequencies above 5 GHz $\psi(f)$ of the models containing $LiNbO_3$ crystal, nonlinearly increases with frequency. The measured dependence $\psi(f)$ was used to calculate the group delay $\tau_g = -\partial(\psi(f))/2\pi\partial f$. Narrow frequency bands with negative values of $\tau_g$ were found (Figure 5).

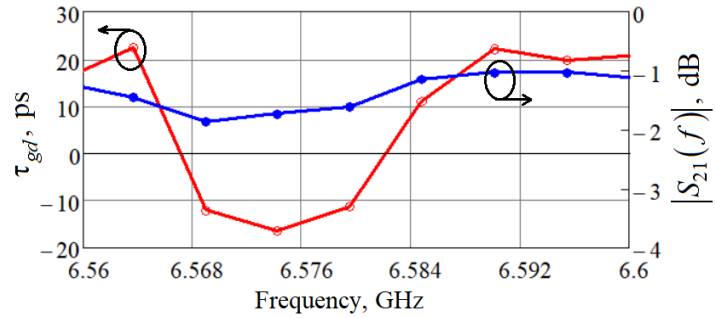

**Figure 5.** Frequency dependence of the group delay time $\tau_{gd}$ and modulus of the transfer coefficient $|S_{21}(f)|$ in the frequency range with negative $\tau_{gd}$.

The considered strip waveguides including bulk lithium niobate crystals can find application for designing quasi-chaotic oscillation shapers in the microwave range.

## 4. Conclusions

The obtained results can be used for the design of quasi-chaotic oscillation shapers in the microwave range. The confirmed occurrence of negative $\tau_{gd}$ is explained by the interference of a large number of normal waves and, due to the difference in their phase velocities, the occurrence of backward motion of the phase front. The appearance of negative $\tau_{gd}$ can be represented as a property of the metamaterial element in the form of a segment of coplanar strip line with partial crystal filling. The use of such an element makes it possible to compensate the distortion of the signal passing through devices with opposite frequency dependence $\tau_{gd}$. However, the practical use of this effect presents certain difficulties due to the narrow frequency range.

## 5. Patents

Patent Ru 2803456. Microwave quasi-chaotic signal shaper module", dated 2022-09-12.

A.G.L. and A.A.A.; writing—original draft preparation, N.D.M., A.A.A. and A.G.L.; writing—review and editing, N.D.M.; visualization, G.A.M.; supervision, A.G.L.; project administration, N.D.M.; funding acquisition, A.G.L. All authors have read and agreed to the published version of the manuscript.

**Funding:** The study was made with the financial support of the Ministry of Science and Higher Education of the Russian Federation within the framework of project No. FEWM-2023-0014.

**Data Availability Statement:** The data presented in this study are available upon reasonable request.

**Acknowledgments:** The authors are grateful to the staff of the collective use center "Impulse" and the scientific and educational center "Nanotechnologies" for their assistance in carrying out measurements.

**Conflicts of Interest:** The authors declare no conflict of interest.